\documentclass[a4paper,twocolumn, a4paper,twocolumn, accepted=2023-12-21]{quantumarticle}
\pdfoutput=1

\usepackage[T1]{fontenc}
\usepackage{amsmath}
\usepackage{hyperref}

\usepackage{amsmath}
\usepackage{relsize}
\usepackage{graphicx}
\usepackage[numbers]{natbib}
\usepackage{dcolumn}
\usepackage{xcolor}
\usepackage{bm}
\usepackage[utf8]{inputenc}
\usepackage[english]{babel}

\newcommand{\TOCITE}{{\color{red} [?] }}

\newcommand{\braket}[1]{\langle #1 \rangle}

\newcommand{\comment}[1]{}

\renewcommand{\d}{{\rm d}}
\newcommand{\Ai}{{\rm Ai}}

\newcommand{\bra}[1]{\langle #1 |}
\newcommand{\ket}[1]{| #1 \rangle}

\newcommand{\BEA}{\begin{eqnarray}}
\newcommand{\EEA}{\end{eqnarray}}
\def\BES#1\EES{\begin{equation}\begin{split}#1\end{split}\end{equation}}

\begin{document}

\title{Energy densities in quantum mechanics}
\author{V. Stepanyan$^{1}$ and A.E. Allahverdyan$^{1,2}$}
\affiliation{
    $^{1}$ Institute of Physics, Yerevan State University, 0025 Yerevan, Armenia\\
    $^{2}$Alikhanian National Laboratory, 0036 Yerevan, Armenia
}

\begin{abstract}
     Quantum mechanics does not provide any ready recipe for defining energy density in space, since the energy and coordinate do not commute. To find a well-motivated energy density, we start from a possibly fundamental, relativistic description for a spin-$\frac{1}{2}$ particle: Dirac's equation. Employing its energy-momentum tensor and going to the non-relativistic limit we find a locally conserved non-relativistic energy density that is defined via the Terletsky-Margenau-Hill quasiprobability (which is hence selected among other options). It coincides with the weak value of energy, and also with the hydrodynamic energy in the Madelung representation of quantum dynamics, which includes the quantum potential. Moreover, we find a new form of spin-related energy that is finite in the non-relativistic limit, emerges from the rest energy, and is (separately) locally conserved, though it does not contribute to the global energy budget. This form of energy has a holographic character, i.e., its value for a given volume is expressed via the surface of this volume. Our results apply to situations where local energy representation is essential; e.g. we show that the energy transfer velocity for a large class of free wave-packets (including Gaussian and Airy wave-packets) is larger than its group (i.e. coordinate-transfer) velocity.
\end{abstract}

\comment{100 words:

Defining a conserving energy density in quantum mechanics is
challenging due to non-commutativity. To solve this problem, we start
from Dirac's equation, where the relativisitic covariance dictates a
unique energy density with a well-defined non-relativistic limit. We
uncover a new form of locally-conserving, spin-dependent, holographic energy that
emerges from the rest-energy. It does not contribute to the global
energy budget but is relevant locally since its density is finite in
the non-relativistic limit. Once the problem is solved we address new
issues and show that for Gaussian and Airy wave-packets the energy
flows at a higher speed than the coordinate.

}

\maketitle

\section{Introduction}

The time-dependent Schroedinger equation provides the locally conserved coordinate density and the corresponding current \cite{landau}. The very understanding of quantum mechanics is based on this picture. It will be useful to have its energy analog: a conserved energy density and the corresponding energy current. They would describe how energy moves and distributes in
space and time. They would also describe the energy content of non-normalizable states. Such states are employed in quantum mechanics \cite{landau,berry}, but the mean energy for them is not defined. 
In classical statistical mechanics, the energy density and current are defined from the joint probability of coordinate and momentum. This method does not work in the quantum situation, since the kinetic energy and coordinate do not commute. This may be solved via a
quasiprobability for the kinetic energy and coordinate \cite{cohen}, but it is unclear which one to select 
due to the variety of quasiprobabilities and their results.

We start from a possibly fundamental description that underlies
the non-relativistic quantum mechanics, {\it viz.} Dirac's equation
for relativistic covariant bispinor field
\cite{davydov,bere,thaller}. It has an energy-momentum tensor
that contains among its components a unique energy density \cite{bere}. 
As seen below, the energy density is unique for two reasons. First, it follows 
from the unique relativistically invariant Lagrangian. In contrast, 
non-relativistic Lagrangian is not unique and different Lagrangians produce different results for 
non-relativistic energy current. Second, the energy density is immune to the known dilemma of 
symmetric {\it versus} anti-symmetric energy-momentum tensor \cite{landau}. 

Dirac's equation is a cornerstone of quantum field theory, but the time-dependent dynamics of the
single-particle Dirac's equation is still controversial \cite{davydov,thaller,bere}. It is unclear what the definition of the operational coordinate is, how to deal with negative energies, {\it etc}. These issues disappear in the non-relativistic limit. This limit employs the power of relativistic field theory with its important predictions (spin, antiparticles, {\it etc}) \cite{davydov,bere,thaller}, but avoids problems related to the relativistic time-space description. 

We show that in the non-relativistic limit Dirac's energy density naturally separates into the non-relativistic energy and the rest energy. The first of them is based on the Terletsky-Margenau-Hill quasiprobability, and also coincides with Madelung's energy density that includes the quantum potential. It correctly recovers the full (mean) energy of the quantum particle. In energy eigenstates with wave-function $\phi_n({\bf r})$ and energy $E_n$, this expression predicts energy density $E_n|\phi_n({\bf r})|^2$. 

We applied this non-relativistic energy density for deducing the energy transfer velocity of a wave-packet and showed that for Gaussian (and also Airy) packets this velocity is generically larger than the coordinate transfer velocity. Energy transfer velocity cannot be studied without defining the energy density, but several conclusions on this velocity are independent from details of this definition. 

An interesting feature of the non-relativistic energy density is that for free motion it does assume negative values. The overall energy is still positive, but lower than the local rest energy. We found states, where this negativity takes place for finite times only. We conjecture that normalized pure states cannot provide non-negative energy density for all times and coordinates. 

There is a fine-grained structure to the rest energy, since it can be separated into the bulk contribution and an additional non-relativistic, spin-dependent part that is locally conserved. This new form of energy has a holographic character, since it is a divergence of a local vector; i.e. its content in a volume $V$ is expressed via the surface integral over $\partial V$. It nullifies for finite-motion stationary states (without magnetic field), but is not zero already for stationary states that describe scattering. The full space-integral of this new energy density is zero, i.e. it does not contribute to the total energy balance. We shall illustrate this new form of energy via examples.


This paper is organized as follows. The next section reviews the problem of energy density in quantum mechanics and discusses several intuitive conditions to be expected from this quantity. Section \ref{density} recalls the energy density and current deduced from Dirac's equation. Section \ref{limit} describes the non-relativistic limit of the energy density. This section also makes connection with the hydrodynamic approach and discusses the negativity of the energy density, Section \ref{holographic} shows that there is an additional, holographic, locally-conserving energy that comes from the rest-energy; its features are exemplified in section \ref{stato}. Section \ref{energy-transfer} shows how to define the energy transfer velocity and discusses some of its features for Gaussian and Airy wave-packets. We briefly summarize in the last section. 

Appendices contain several relevant messages, as well as some derivations. Appendix \ref{aa1} recalls how expressions (\ref{vava}, \ref{vovo}) for the energy density and current are derived from the unique relativistic invariant Langrangian and the energy-momentum tensor. Appendix \ref{aa2} studies the derivation of the energy density (\ref{gru}) from a non-relativistic Lagrangian. This Lagrangian is not unique, and the two main choices are presented in detail. Appendix \ref{aa3} looks at 1d Gaussian wave-packets, their energy density and energy transfer speed comparing it with the coordinate transfer speed. Appendix \ref{aa4} does the same for Airy wave-packets and also explains the physical meaning of these packets via the energy density (\ref{gru}). Eq.~(\ref{rable}) in Appendix \ref{aa3} provides an example of 1d normalizable state, whose energy density $\rho(\bm r, t)$ is negative for a finite ranges of $\bm r$ and $t$ only. Appendix \ref{simon} demonstrates for certain states the energy transfer velocity can be slower than the coordinate transfer velocity.
Appendix \ref{resto} discusses the positivity feature of the total energy density. 
Appendix \ref{aa5} calculates the holographic energy (\ref{kakav}) for Landau levels (free 2d electrons in a magnetic field) that are basic for the modern solid state physics.

\section{Problem definition}
\label{related}

The problem of energy density is mainly in the overabundance of possible functions that could be called energy density. 
One way to understand this overabundance is to note that the definition of energy density in classical mechanics is unique, because there is a unique joint density for the coordinate and and momentum, and the energy density can be obtained as the local value deduced from this density \cite{cohen1979,cohen1984}. Now in quantum mechanics, the above joint density is to be replaced by a quasi-probability that is essentially not unique. 

Some general energy density families are derived in \cite{cohen1979,cohen1984,juan10}. The four most common energy densities (out of potentially infinite number of them) are presented in \cite{mathews_dens,cohen1979,cohen1984}, however, the choice between them is unclear. Each of these energy densities satisfies different properties that may seem to be useful for concrete problems. For example, some of these functions have been connected to heat and energy transfer, work, electron orbitals, quantum arrival times etc \cite{muga05, Wu_2009, tsirelson16, sanchez14, geraldine17, tachibana01}. Let us list some necessary and optional properties of an energy density function.

$\bullet$ Energy density $\rho_\circ(\bm r,t)$ must be quadratic from the wave function $\phi(\bm r,t)$ (or linear over the density matrix) to preserve locality. Put differently, the marginalized density should refer to a quantum subsystem. 

$\bullet$ Energy density $\rho_\circ(\bm r,t)$ must be locally conserved 
\BEA 
\label{g1}
\dot{\rho}_\circ(\bm r,t) + \bm \nabla \bm J_\circ(\bm r,t) = \phi^\dagger(\bm r,t) \phi(\bm r,t)\, \partial_t U(\bm r,t),
\EEA
where $U(\bm r,t)$ is the potential energy that can generally depend on time due to external sources, $\bm \nabla=\partial/\partial \bm r={\rm div}$, and where $\phi(\bm r,t)$ is the wave function (generally a spinor). For $\partial_t U=0$ (no time-dependent external fields), (\ref{g1}) means literal local conservation of energy. For $\partial_t U\not =0$ it means that the sources of energy injection (or extraction) act locally.

$\bullet$ The integral of energy density must be equal to the average energy 
\BEA
\label{g2}
\int \rho_\circ(\bm r,t) \,\d  r = \int \phi^\dagger {H}\phi \,\d  r, 
\EEA
where $H$ is the Hamiltonian.
    
$\bullet$ An optional property is for the kinetic term of energy density to be positive 
\BEA
\label{g3}
\rho_\circ(\bm r,t) - U(\bm r,t)\phi^\dagger(\bm r,t)\phi(\bm r,t) \geq 0.
\EEA
A possible rationale of (\ref{g3}) is that the kinetic energy is a positive operator, and hence its density is also postulated to be positive. 
Eq.~(\ref{g3}) is employed when working in electron orbital models \cite{tsirelson16}, for studying energy current in superconductors \cite{canadian}, and for didactic purposes \cite{mita}. However, there are works suggesting that the negativity of kinetic energy density can in fact be useful \cite{berry10, tachibana01}. 

$\bullet$ Another optional property is for the energy density to be of the form 
\BEA
\label{g4}
\rho_\circ = E\phi_E^\dagger\phi_E~~ {\rm when}~~ \mathcal{H}\phi_E = E\phi_E.
\EEA
Eq.~(\ref{g4}) says that for stationary states, where the energy is well-defined, the uncertainty of the energy is to be determined by the uncertainty of the position only. 

Conditions (\ref{g3}) and (\ref{g4}) contradict each other: writing (\ref{g3}) for a stationary state and combining it with (\ref{g4}) we get $(E-U(\bm r))\phi_E^\dagger(\bm r)\phi_E(\bm r) \geq 0$. This is violated precisely when the quantum particle can be in a classical forbidden coordinate domain (this closely relates to quantum tunneling, which is however a non-stationary effect). 

Among various possibilities, the literature focuses on the following two expressions for energy density \cite{mathews_dens}:
\BEA
\label{oliki}
&&\rho=- \frac{\hbar^2}{4m}(\,[\Delta\phi^\dagger]\phi + \phi^\dagger \Delta \phi) 
+ U\phi^\dagger\phi, \\
&&\widetilde{\rho}=\frac{\hbar^2}{2m} {\bm\nabla}\phi^\dagger{\bm\nabla}\phi +U\phi^\dagger\phi.
\label{boliki}
\EEA
Now $\rho$ holds all above conditions besides (\ref{g3}), while $\widetilde{\rho}$ holds all above conditions besides (\ref{g4}).
We show that the absence of (\ref{g3}) is not really a drawback, since within a consistent relativistic derivation it comes out as a part of the full energy which is still positive (due to the rest energy) even if $\rho(\bm r)<0$; see section \ref{negus}.
Both $\rho$ and $\widetilde{\rho}$ are derived from different classical Lagrangians; see Appendix \ref{aa2}. Both of them have Born form, i.e. they are given as ${\rm tr}[R{\cal X}]$, where $R$ is the density matrix of the quantum system, and ${\cal X}$ is an Hermitian operator; cf.~(\ref{klu}, \ref{dag}). Another difference is that $\rho$ contains the kinetic energy operator, while $\widetilde{\rho}$ is composed of two momentum operators; see (\ref{klu}, \ref{dag}). Hence $\widetilde{\rho}$ can apply only to those situations, where the energy is written as the square of a Hermitian operator, while ${\rho}$ applies more generally.

Therefore, a choice must be made when working with energy densities between the properties (\ref{g3}) and (\ref{g4}), or none of them. Instead of making this choice ourselves, we derive the energy density from the possibly fundamental representation of quantum mechanics: Dirac's equation.

{\color{black}
Note that both (\ref{oliki}) and (\ref{boliki}) are obtained as local values of certain quasi-probabilities:
\BEA
\label{olga}
&&\rho(\bm r)=\int \d^3p\, (\frac{\bm p^2}{2m}+U(\bm r))W_{\rm TMH}(\bm r,\bm p),\\
\label{vera}
&&\widetilde{\rho}(\bm r)=\int \d^3p\, (\frac{\bm p^2}{2m}+U(\bm r))\widetilde{W}(\bm r,\bm p),\\
\label{nata}
&& \widetilde{W}(\bm r,\bm p)=2W(\bm r,\bm p)-W_{\rm TMH}(\bm r,\bm p),
\EEA
where (\ref{olga}) with Terletsky-Margenau-Hill quasi-probability follows from (\ref{klu}), while (\ref{vera}, \ref{nata}) follow from (\ref{olga}) and definition of the Wigner quasi-probability $W(\bm r,\bm p)$. A quasi-probability representation is important since it shows that the energy density is meaningful in the semi-classical limit. In addition, both (\ref{oliki}) and (\ref{boliki}) correctly transform during a Galilean boost from one reference frame to another; see Appendix \ref{boost-a}. Both for (\ref{oliki}) and (\ref{boliki}) this transformation relies on the momentum density defined via the Terletsky-Margenau-Hill quasi-probability for coordinate and momentum. 
}

\section{Energy density and current from Dirac's equation} 
\label{density}

Dirac's equation governs bispinor $\psi$ for a relativistic spin-$\frac{1}{2}$ particle \cite{bere,davydov,thaller}:
    \begin{align}
    \label{kur}
& i\hbar\dot{\psi} ={\cal H}\psi\equiv
mc^2\beta\psi+U(\bm{r},t)\psi-i\hbar c(\bm \nabla\bm{\alpha}\psi),\\
& \bm\alpha = \begin{pmatrix}0 & \bm\sigma \\ \bm\sigma & 0\end{pmatrix},~~ 
\beta = \begin{pmatrix}1 & 0 \\ 0 & -1\end{pmatrix}, ~~  
\partial_t{\psi}=\dot{\psi}, ~~
\bm\nabla=\partial_{\bm r},
\label{fantom}
    \end{align}
where $U(\bm{r}, t)$ is a potential energy, $\bm{\alpha}=\{\alpha_i\}_{i=1}^3$ and $\beta$ are $4\times 4$ Dirac's matrices, $\bm{\sigma}=\{\sigma_i\}_{i=1}^3$ are Pauli's matrices. $U(\bm{r},t)$ is externally time-dependent indicating on processes of work-exchange. The local energy conservation, energy density and current read (resp.) \cite{bere,greiner}:
\begin{align}\label{eq:DiracDensityAndFlux}
& \dot\varrho(\bm{r}, t) + \bm \nabla \bm{\mathcal{J}}(\bm{r}, t)-\dot U(\bm{r},t)\psi^\dagger\psi = 0,\\
\label{vava}
& \varrho = \frac{i\hbar}{2}\bigg(\psi^\dagger\dot\psi-\dot\psi^\dagger\psi\bigg) = mc^2\psi^\dagger\beta\psi+U\psi^\dagger\psi + \nonumber\\
&+ \frac{i\hbar c}{2}\bigg[(\bm \nabla\psi^\dagger\bm \alpha)\psi - 
\psi^\dagger(\bm \nabla\bm\alpha\psi)\bigg],\\
&\bm{\mathcal{J}} = \frac{i\hbar c}{2}\bigg(\psi^\dagger\bm\alpha\dot\psi - 
\dot\psi^\dagger\bm\alpha\psi\bigg) =\nonumber Uc\psi^\dagger\bm\alpha\psi+\\&+ \frac{i\hbar c^2}{2}\bigg[(\bm \nabla\psi^\dagger\bm 
\alpha)\bm\alpha\psi-\psi^\dagger\bm\alpha(\bm \nabla\bm\alpha\psi)\bigg],
\label{vovo}
\end{align}
where (\ref{eq:DiracDensityAndFlux}) is the local conservation law with $\dot U(\bm{r},t)\psi^\dagger\psi$ being the local source of work. Eqs.~(\ref{eq:DiracDensityAndFlux}--\ref{vovo}) are derived via the unique, relativistically invariant Lagrangian and the energy-momentum tensor; see Appendix \ref{aa1}. We emphasize that the uniqueness of (\ref{eq:DiracDensityAndFlux}--\ref{vovo}) follows from the relativism, and it is absent in the non-relativistic physics. In particular, it is absent in the non-relativistic Lagrangian formalism; see Appendix \ref{aa2}.

Eq.~(\ref{kur}) can be employed for checking (\ref{eq:DiracDensityAndFlux}--\ref{vovo}). $\varrho$ in (\ref{vava}) relates to the $00$-component of the energy-momentum tensor of Dirac's field \cite{bere}. For relativistic fields with non-zero spin, the definition of the latter tensor is known to be non-unique: the canonic (Noether's) tensor is not
symmetric, and there is symmetric (Belinfante-Rosenfeld) tensor that is
employed in gravitation and that agrees with the canonic one globally,
but not locally \cite{bere}. However, for Dirac's field the two tensors
relate to each other via the symmetrization of indices; hence this
ambiguity of energy-momentum tensors does not affect $\varrho$ and
$\bm{\mathcal{J}}$ (up to a rotor field for the latter) \cite{bere}.
Hence the energy-density for Dirac's field is indeed well-defined in
contrast to e.g. the momentum density. 

Eqs.~(\ref{kur}, \ref{vava}) imply expectedly for the mean energy:
\BEA
\label{daad}
\int\d^3 r\varrho(\bm r,t)=\int\d^3 r\psi^\dagger (\bm r,t) {\cal H}\psi (\bm r,t). 
\EEA
The conservation of density $\psi^\dagger\psi$ reads from (\ref{kur}) \cite{bere,davydov,thaller}:
\BEA
\label{den}
\partial_t(\psi^\dagger\psi)+\bm{\nabla}[c\psi^\dagger\bm{\alpha}\psi]=0.
\EEA
Note that for $U=0$ the density $\psi^\dagger\psi$ is non-negative, in constrast to the energy density 
$\varrho$ in (\ref{vava}) that can hold $\varrho(\bm r)<0$ for some $\bm r$; e.g. because $\beta$ in (\ref{fantom}) has a negative eigenvalue \cite{davydov}. This is an aspect of the spin-statistics theorem: half-integer spin wave-equations lead to non-negative density, but not a non-negative energy density. For integer spins the situation is opposite: the energy density is non-negative, while the density is not \cite{davydov}. 

\section{Energy density in the non-relativistic limit} 
\label{limit}

\subsection{Derivation}

The introduction of this limit in (\ref{kur}) starts with representing the bispinor $\psi$ via two spinors $\varphi$ and $\chi$ and introducing a phase-factor with the rest energy $mc^2$ \cite{bere,greiner}:
    \begin{equation}
    \label{gro}
        \psi = \begin{bmatrix}\varphi \\ \chi\end{bmatrix} e^{-\frac{i}{\hbar}mc^2t},
    \end{equation}
Eqs.~(\ref{kur}, \ref{gro}) lead to differential equations
for two spinors $\varphi$ and $\chi$: 
\begin{align}
\label{kora}
&i\hbar\dot\varphi=U\varphi-i\hbar c\bm\sigma\bm\nabla\chi,\\
&i\hbar\dot\chi =-2mc^2\chi +U\chi-i\hbar c\bm\sigma\bm\nabla\varphi.
\label{kamilla}
\end{align}
Solving (\ref{kora}) perturbatively for a large $c$ we find (\ref{arlanda}), while putting the result into (\ref{kamilla}) we get (\ref{susanna}):
\begin{align}
            \label{arlanda}
            \chi &= -\frac{i\hbar\bm\nabla\bm\sigma\varphi}{2mc} - \frac{i\hbar U\bm\nabla\bm\sigma\varphi}{4m^2c^3} - \frac{\hbar^2\bm\nabla\bm\sigma\dot\varphi}{4m^2c^3}\\
            \label{gla}
               & = -\frac{i\hbar}{2mc}\bm \nabla\bm\sigma\phi + \mathcal{O}(c^{-3}),\\
            \dot\varphi& = \frac{i\hbar}{2m}\Delta\varphi - \frac{iU}{\hbar}\varphi + \frac{i\hbar}{4m^2c^2}\bm\nabla\bm\sigma\big( U(\bm\nabla\bm\sigma)\varphi\big)+\\&+ \frac{\hbar^2\Delta\dot\varphi}{4m^2c^2},
        \label{susanna}
\end{align}
where in (\ref{susanna}) we employed $(\bm\nabla\bm\sigma)^2=\Delta$.
Now take $\phi = \varphi + O(c^{-2})$ in (\ref{susanna}) and note that for 
$c\to\infty$ the Schroedinger equation holds for $\phi$ with Hamiltonian $H$:
\BEA
\label{er}
i\hbar \dot\phi =H\phi,\quad H\equiv -\frac{\hbar^2}{2m}\Delta+U.
\EEA
The essence of the non-relativistic limit is that $\phi\gg \chi$. The main difference with the usual way of taking this limit is that we need to retain $\chi$, since (\ref{eq:DiracDensityAndFlux}) has terms $mc^2\chi^\dagger\chi$ that survive in the non-relativistic limit. Using (\ref{gla}, \ref{eq:DiracDensityAndFlux}, \ref{er}) we get
\BEA
\label{eq:DiracDensityCalcd}
&\varrho = \rho+ mc^2(\varphi^\dagger\varphi+\chi^\dagger\chi)  + \mathcal{O}(c^{-2}),\\
&\rho=- \frac{\hbar^2}{4m}(\,[\Delta\phi^\dagger]\phi + \phi^\dagger \Delta \phi) 
+ U\phi^\dagger\phi.
\label{gru}
\EEA
Since $\varphi^\dagger\varphi+\chi^\dagger\chi$ is the particle density [cf.~(\ref{gro}, \ref{daad})], the energy density $\varrho$ in (\ref{eq:DiracDensityCalcd}) separates into the non-relativistic energy density $\rho$ (\ref{gru}) and the rest energy. Now $\rho$ is locally conserved [cf.~(\ref{g1})] as deduced from 
(\ref{er}): 
\begin{gather}
\label{ord}
\dot \rho + \bm \nabla \bm J -\dot U(\bm{r},t)\phi^\dagger\phi= 0,\\
\label{klu}
{\rho}={\rm tr}\Big[R\,\{H, \,\ket{\bm{r}}\bra{\bm{r}}\} \Big],\\
{\bm J}=\frac{\hbar^2}{2m}\Re\Big[ \phi^\dagger\bm\nabla\dot \phi -[\bm\nabla\phi^\dagger]\dot\phi    \Big],
\label{bonbon}
\end{gather}
where $\dot U(\bm{r},t)\phi^\dagger\phi$ in (\ref{ord}) is the work term, $R =\ket{\bm \phi} \bra{\bm \phi}$ is the density matrix, and $\{a,b\}=\frac{1}{2}(ab+ba)$ is the (half) anticommutator. 
Eq.~(\ref{gru}) implies for the mean non-relativistic energy [cf.~(\ref{g2}, \ref{daad})]:
\BEA
\label{mean}
\int\d^3 r\,\rho(\bm{r},t)=\int\d^3 r\,\phi^\dagger(\bm r,t) H\phi (\bm r,t),
\EEA
where $\int\d^3 r\,\phi^\dagger(\bm{r}, t)\phi(\bm{r}, t)=1$. 
For stationary states of (\ref{er}) the spin and coordinate factorize, $\phi_s=|s\rangle {\phi}(\bm{r})$, where $H{\phi}(\bm{r})=E_n{\phi}(\bm{r})$. Now $\rho$ reduces to the particle's density [cf.~(\ref{g4})]
\BEA
\label{doors}
\rho(\bm{r})=E_n|\phi_n({\bf r})|^2. 
\EEA
Moreover, $\bm J=0$ whenever $\phi_n({\bf r})=\phi_n^{*}({\bf r})$, i.e. the energy does not flow in stationary states with finite motion.

Eqs.~(\ref{klu}, \ref{bonbon}) show that $\rho$ and $\bm J$ relate to the Terletsy-Margenau-Hill quasiprobability \cite{kirkwood,terletsky,dirac,barut,margenau,armen,matteo,hofer_quasi,lobejko_quasi,francica_quasi}. $\rho$ is interpreted as the joint probability of energy and coordinate and $\bm J$ as the joint probability of energy and probability current. The usage of this quasiprobability for the local energy was postulated in \cite{mclennan,hardy}. In particular, Ref.~\cite{hardy} applied this energy current to phonon thermal conductivity in solids. Here we derived this postulate from Dirac's equation. Eq.~(\ref{klu}) is also found if assuming two non-interacting particles in a state $\phi(\bm{r}_1,\bm{r}_2)$, we trace out the second particle, and denote by $R$ the density matrix of the first particle.

Appendix \ref{aa2} explains that (\ref{gru}, \ref{bonbon}) can be derived via a Lagrangian for (\ref{er}). The choice of the non-relativistic Lagrangian is not unique in contrast to the relativistic situation. The non-relativistic Lagrangian that produces (\ref{gru}, \ref{bonbon}) is unusual, since contains second-order space-derivatives. 

\subsection{Relations with the hydrodynamic approach}

Eq.~(\ref{gru}) coincides with the prediction of the quantum hydrodynamic approach \cite{madelung,takabayasi}.
Define $\phi_s(\bm r,t)=|\phi_s(\bm r,t)|e^{i\Gamma_s(\bm r,t)/\hbar}$, where $s=1,2$ refers to the spinor components. Now the quantum dynamics reduces to classical hydrodynamics (in Euler's picture), for two fluids ($s=1,2$), where $\bm v_s(\bm r,t)=\frac{\bm \nabla \Gamma_s(\bm r,t)}{m}$ is the local velocity, while $|\phi_s(\bm r,t)|^2$ is the local density \cite{takabayasi}. The energy density of this hydrodynamical system, 
\begin{equation}
{\sum}_{s=1}^2 |\phi_s(\bm r,t)|^2 \bigg[\frac{m\bm v_s^2(\bm r,t)}{2}+U(\bm r,t) - \frac{\hbar^2}{2m}\frac{\Delta {|\phi_s|}}{|\phi_s|} \bigg],
\label{hydro}
\end{equation}
coincides with $\rho (\bm r,t)$ in (\ref{gru}, \ref{klu}). In (\ref{hydro}) we have kinetic energy, potential energy, and Bohm's (quantum) potential \cite{takabayasi} that comes from the kinetic energy in (\ref{gru}). Recall that within the hydrodynamic approach, Bohm's potential comes from a pressure gradient. Eq.~(\ref{klu}) is more general than (\ref{hydro}) since it applies to mixed states. Note that (\ref{klu}) contains the weak value of the kinetic energy \cite{cohen,aharonov,pop}. 

\subsection{Negativity of non-relativistic energy density  }
\label{negus}

For simplicity, we focus here on the free motion ($U=0$), because the issues of negativity are relevant already here. 
It should be obvious from (\ref{gru}) that the energy density $\rho(\bm r,t)$ can be negative for certain $\bm r$ and $t$ despite of the positive mean energy (\ref{mean}); cf.~(\ref{g3}, \ref{g4}). In Appendix \ref{aa3} we work out the free ($U=0$) motion of a Gaussian wave-packet and demonstrate the above negativity explicitly; see Eq.~(\ref{rable}) and the discussion after it. This negativity is a quantum effect. Its origin can be traced back to the  Terletsky-Margenau-Hill quasiprobability that also needs to be sometimes negative due to non-commutativity of coordinate and momentum \cite{armen}.   

The negativity of $\rho(\bm r,t)$ does not mean that the full energy density $\varrho(\bm r,t)$ is negative. Indeed, according to (\ref{eq:DiracDensityCalcd}), $\varrho(\bm r,t)$ contains (besides $\rho(\bm r,t)$) the contribution from the rest energy; cf.~(\ref{susanna}, \ref{eq:DiracDensityCalcd}) and recall that $\varphi=\phi+{\cal O}(c^{-2})$. The rest energy is positive and large for each $\bm r$, where $\phi(\bm r)\not =0$. Appendix \ref{resto} shows that in the proper non-relativistic limit -- where the characteristic distances are much larger than the Compton length -- the rest energy is the main contribution in $\varrho>0$. Appendix \ref{resto} also discusses the situation with wave-function zeros, $\phi(\bm r_0) =0$, and shows that whenever a zero holds additionally $\bm\nabla\phi(\bm r_0)\not =0$, we still get $\varrho(\bm r)\geq 0$ for $\bm r\approx\bm r_0$. This is due to factor $mc^2\chi^\dagger\chi$ in (\ref{eq:DiracDensityAndFlux}).

Once the total energy density is non-negative, $\varrho(\bm r,t)\geq 0$, the negativity of $\rho(\bm r,t)$ is not a physical drawback. In particular, this negativity itself cannot serve as an argument for selecting the alternative kinetic energy density $\widetilde{\rho}$ that is non-negative by definition; see (\ref{boliki}, \ref{porto}). Recall that $\widetilde{\rho}$ is deduced from a non-relativistic Lagrangian, and not from any relativistic treatment; see Appendix \ref{aa2}. 

Returning to $\rho(\bm r,t)<0$ for certain values of $(\bm r,t)$, it is interesting to understand whether the negativity of $\rho(\bm r,t)$ is inevitable over time if it was absent initially. When asking this question we shall concentrate (for simplicity and clarity) on the free particle case: $U=0$. First of all, we note that it is not difficult to construct non-normalizable (non-stationary) states, where $\rho(\bm r,t)\geq 0$ for all $(\bm r,t)$. The simplest example of this is provided by a superposition of two plane waves: 
\begin{align}
\phi(\bm r,t)&=e^{i\bm k_1\bm r-\frac{i\hbar t\,\bm k_1^2}{2m}}+e^{i\bm k_2\bm r-\frac{i\hbar t\,\bm k_2^2}{2m}},\\
\rho(\bm r,t)&=\frac{\hbar^2(k_1^2+k_2^2)}{2m}\times\\&\times\bigg[1+\cos[\bm r(\bm k_1-\bm k_2)+\frac{t(k_2^2-k_1^2)}{2m\hbar}  ]\bigg],\nonumber 
\end{align}
where $\bm k_1$ and $\bm k_2$ are two wave-vectors, and $k_1=|\bm k_1|$, $k_2=|\bm k_2|$. 
It is seen that $\rho(\bm r,t)\geq 0$. 

For normalizable states, we searched for several classes of states and did not find states for which 
$\rho(\bm r,t) \geq 0$ for all values of $\bm r$ and all values of $t$ (both negative and positive). In particular, this holds 
for Gaussian wave-packets (pure state), though there are examples of such states, where $\rho(\bm r,t) < 0$ takes place
for a finite range of $t$ only; see Appendix \ref{aa3}. We conjecture that there are no normalizable states for which $\rho(\bm r,t)\geq 0$ for all $\bm r$ and all $t$ ($t<0$ and $t>0$). 

There is a remote analogy between $\rho(\bm r,t)<0$ and the fact that post-renormalized energy densities of certain quantum field theories show negative values; see \cite{lectures,ford,ford_chin,ann} for reviews. There the negativity is real, i.e. it concerns the full energy density because according to the current understanding of the quantum field theory, the infinities omitted during some schemes of renormalization are not real energies. Once the negativity is real it would create problems for gravitation. This fact led people to look for inequalities for bounding this negativity \cite{lectures,ford,ford_chin,ann}. Interestingly, these inequalities also found applications for bounding the negative values of $\rho(\bm r,t)$ \cite{ann}. We tried to apply the results of Ref.~\cite{ann} for studying the above conjecture (which is about the emergence of negativity and not so much about its magnitude), but so far without success.  

\section{Energy transfer velocity} 
\label{energy-transfer}

\subsection{Energy transfer velocity defined via the space-integrated energy current} 
\label{macro-energy-transfer}

The following important question cannot be addressed without the notion of quantum mechanical energy density, though (as we shall see) the answer to an extent does not depend on the details of the definition. What is the energy transfer velocity for a freely propagating ($U(\bm r,t)=0$) quantum wave packet, and how does it differ from the velocity of the coordinate transfer? This question is relevant for all quantum fields dealing with excitation transfer, but it does not seem to be addressed so far. It has a remote analog in optics, where the energy transfer velocity (group velocity) is compared with the unphysical phase-velocity \cite{bril,milo}. In our situations both velocities are physical, and the group velocity refers to the coordinate transfer.  

We start with a wave packet with 1d Schroedinger wave-function $\phi(x,t)$, energy density (\ref{gru}) and energy current (\ref{bonbon}). The spin degree of freedom is irrelevant provided that its wave-function factorizes  $\phi_s(x,t)=|s\rangle \phi(x,t)$; see (\ref{gru}, \ref{bonbon}). Looking at the local conservation 
\begin{equation}
\partial_t(\phi^*\phi)+\partial_xj=0,~ j(x,t)=\frac{i\hbar}{2m}(\partial_x\phi^*\phi-
\phi^*\partial_x\phi),
\end{equation}
of density, we define the coordinate transfer velocity as 
\begin{align}
& v_{\rm cor}(t)=\int\d x\, j(x,t)=\frac{1}{m}\int\d x\, \phi(x,t)P\phi(x,t)\nonumber\\
& =({1}/{m})\braket{P}(t),~~~ P=({i}/{\hbar})\partial_x.
\label{oswald}
\end{align}
$v_{\rm cor}(t)$ relates to the mean momentum, which is also the mean group velocity \cite{bril,milo}; cf.~(\ref{hydro}). Likewise, a sensible (though by no means exclusive) definition of the energy transfer velocity will be 
\BEA
v_{\rm en}(t)=\frac{\int\d x\, J(x,t)}{\int\d x\, \rho(x,t)}=\frac{\braket{P^3}}{m \braket{P^2}},
\label{prangle}
\EEA
where we used (\ref{gru}, \ref{bonbon}, \ref{mean}) and employed 
\BEA
\label{brave}
\int\d x\, J(x,t)=\frac{1}{2m^2}\braket{P^3}.
\EEA
This intuitive relation should hold for any sensible definition of the non-relativistic energy density, and not only (\ref{gru}, \ref{bonbon}). (Eq.~(\ref{mean}) has the same general status.) Altogether, (\ref{prangle}, \ref{brave}) should hold for any definition of energy density and its current. In particular, it holds for the definition discussed around (\ref{lagr3}). {\color{black} Note that (\ref{brave}) has a clear classical meaning: formally the same expression can be derived for a free classical particle, whose coordinate-momentum probability density function holds the Liouville equation. }

We assume that both coordinate and energy move in the same direction:
\BEA
\label{komma}
\braket{P^3}>0,\qquad \braket{P}>0.
\EEA
Eqs.~(\ref{oswald}--\ref{brave}) imply:
\BEA
v_{\rm en}-v_{\rm cor}=\frac{\braket{P^3}-\braket{P}\braket{P^2}}{m\braket{P^2} }.
\label{praz}
\EEA
Eq.~(\ref{praz}) is positive for at least two broad classes of probability densities $f(P)$ of $P$. First, consider
\BEA
\label{graz}
f(P)=g(P-\braket{P}), \qquad g(-P)=g(P),
\EEA
where $g(P)$ is a symmetric probability density, $\int\d Pg(P)=1$, and where $\braket{P}=\int\d Pf(P)$. 
This class includes Gaussian wave packets, by far the most frequently studied and realized example of free-particle motion. 
It is easy to derive from (\ref{praz}, \ref{graz}) that $\braket{P^3}>0$ holds due to $\braket{P}>0$, and that:
\BEA
v_{\rm en}-v_{\rm cor}=\frac{2(\braket{P^2}-\braket{P}^2)\braket{P}}{m\braket{P^2} }\geq 0,
\label{raz}
\EEA
i.e. energy is transferred at a larger velocity. Another class of states is those with a non-negative random variable $P$, i.e. $f(P<0)=0$. Now 
\begin{equation}
v_{\rm en}-v_{\rm cor}=\frac{\Big \langle\, \Big(\braket{P}-\braket{P}\Big)\Big(P^2-\braket{P^2}\Big)\,\Big\rangle}
{m\braket{P^2} }\geq 0,
\label{razraz}
\end{equation}
follows from the fact that $P^2$ is a monotonically increasing function of $P$ for $P>0$. 

There are however also cases, where (\ref{komma}) holds, but $v_{\rm en}-v_{\rm cor}<0$; see Appendix \ref{simon}. To our present understanding, these are isolated cases, which refer to transient motion and do not correspond to any known wave packet; more work is needed to understand this point. 

\subsection{Energy transfer velocity defined via the most likely values} 

Airy wave packets is an interesting example of quantum states that move without dispersion (i.e. as a whole) and with acceleration \cite{berry,observation}; see Appendix \ref{aa4}. However, these packets cannot be normalized, i.e. the definition (\ref{prangle}) of energy transfer velocity is not applicable. More fine-grained definitions are needed for such cases. 

Appendix \ref{aa3} shows the same message $v_{\rm en}(t)\geq v_{\rm cor}(t)$ -- for both Airy and Gaussian wave packets -- within another definition of the energy transfer velocity, where analogs of $v_{\rm en}(t)$ and $v_{\rm cor}(t)$ are defined via the velocity of peaks (most probable values) for $\rho(x,t)$ and $\phi^*(x,t) \phi(x,t)$, respectively. This (second) expression of energy transfer velocity already depends on the employed definition (\ref{gru}). Appendix \ref{aa4} also shows that the energy density (\ref{gru}) does explain the physical meaning of Airy wave-packets.

\section{Rest-mass energy and holographic energy} 
\label{holographic}

Returning to (\ref{eq:DiracDensityCalcd}), note that the particle density $\varphi^\dagger\varphi+\chi^\dagger\chi$ is also conserved locally; cf.~(\ref{den}). Moreover, using $\sigma_i\sigma_j=\delta_{ij}+i\epsilon_{ijk} \sigma_k$ in  (\ref{gla}), we note that in the non-relativistic limit the rest energy (\ref{gru}) can be written as a sum of two quantities that are  conserved separately:
\BEA
\label{gog}
&mc^2(\varphi^\dagger\varphi +\chi^\dagger\chi)= mc^2n_0+\rho_s,\\
\label{kakav}
&\rho_s=- \frac{i\hbar^2}{4m} {\bm\nabla}\phi^\dagger\cdot \bm{\sigma}\times \bm{\nabla}\phi,\\
&n_0=\varphi^\dagger\varphi + \frac{\hbar^2}{4m^2c^2}
\bm\nabla\phi^\dagger\cdot\bm\nabla\phi \geq 0. 
\label{pal}
\EEA
$\rho_s$ is a ${\cal O}(c^{-2})$ part of the particle's density that leads to a non-relativistic, and locally conserved energy density. The local conservation of $\rho_s$ can be deduced from (\ref{er}). Alternatively, we note that $\rho_s$ is a divergence of a vector field $\bm\Upsilon(\bm r,t)$:
\begin{equation}
\label{len}
\rho_s=\bm\nabla\bm\Upsilon,~~~~ \dot \rho_s + \bm \nabla J_s = 0, ~~~~ \bm J_s=-\dot{\bm{\Upsilon}},
\end{equation}
\begin{align}
&\bm\Upsilon=\frac{\hbar}{4m}\Re[\phi^\dagger\bm\sigma\times\bm P\phi]\nonumber\\
&=\frac{\hbar}{4m}\Re\,{\rm tr}\Big[\ket{\bm{r}}\bra{\bm{r}}\bm\sigma\times\bm{P} R\Big],
\label{sta}
\end{align}
where $R$ is the density matrix [cf.~(\ref{klu})], and $\bm P=-i\hbar\bm\nabla$ is the momentum operator. Note that the current $\bm J_s$ is expressed via the time-derivative of $\bm\Upsilon$. Eq.~(\ref{len}) means that $\rho_s$ does not have a global content [cf.~(\ref{kakav})]:
\begin{equation}
\int\d r^3\rho_s(\bm r,t)=0~~ {\rm for}~~ \int\d r^3\phi^\dagger(\bm r,t)\phi(\bm r,t)=1.
\label{zero}
\end{equation}
{\color{black} Now (\ref{sta}) is the Terletsky-Margenau-Hill quasi-probability for the coordinate and operator $\bm\sigma\times\bm{P}$. Hence the non-relativistic energy is expressed via the Terletsky-Margenau-Hill quasi-probability [see (\ref{klu})], while for $\rho_s=\bm\nabla\bm\Upsilon$, it is $\Upsilon$ that is expressed via the suitable Terletsky-Margenau-Hill quasi-probability. }

Eq.~(\ref{len}) also shows that the content $\int_V\d^3 r\rho_s$ of $\rho_s$ in a finite volume $V$ is expressed as the integral of $\bm\Upsilon$ over the boundary $\partial V$; i.e. $\rho_s$ is a holographic quantity. Now $\bm\Upsilon$ is essentially spin-dependent and hence quantum, as witnessed by the $\hbar$ factor in (\ref{sta}). For a finite-motion stationary state (without magnetic field) we get $\rho_s=0$, since spin and coordinate factorize, $\phi_s(\bm r)=|s\rangle\phi(\bm r)$, and $\phi(\bm r)=\phi^*(\bm r)$; see (\ref{kakav}). 

\comment{\BEA
& \dot \rho_s + \bm \nabla J_s = 0, \qquad
\bm J_s = \frac{\hbar U(\bm{r})}{4m}\bm \nabla \times (\phi^\dagger\bm\sigma\phi) \\
& + \frac{\hbar^3}{8m^2}\epsilon_{ijk}\sigma_j(\bm\nabla\partial_i\phi^\dagger\partial_k\phi-
     \partial_i\phi^\dagger\bm\nabla\partial_k\phi).
\EEA}

$\rho_s$ in (\ref{kakav}) is invariant with respect to space-inversion $\bm r\to-\bm r$, since $\bm \Upsilon$ in (\ref{sta}) is vector product of vector and pseudo-vector (hence $\Upsilon$ is vector), while the non-relativistic spinor under $\bm r\to-\bm r$ is just multiplied by a phase factor \cite{bere} that disappears from (\ref{kakav}). In the context of (\ref{sta}) recall that the quantity $\bm \sigma\cdot\bm P$ is a pseudo-vector and refers to particle's helicity \cite{bere}.

$\rho_s$ is also invariant with respect to time-inversion, as any energy should be \cite{davydov,landau}. Recall that the time-inversion of a spinor $\phi$ is defined as \cite{davydov,landau}
\BEA
\phi^{[T]}=\hat\pi \phi^*, \qquad
\hat\pi=\begin{pmatrix} 
0 & -1 \\
1 & 0 \end{pmatrix}, 
\label{ti}
\EEA
where $\phi^*$ means complex conjugation. Eq.~(\ref{ti}) means that besides the complex conjugation, which is related to time-invertion of a wave function, the spinor should be subject to an additional unitary transformation $\hat\pi$, which reflects the fact that the eigenvalues of the spin itself change sign under time-inversion \cite{landau}. We now employ $\bm\sigma^{*}=\hat\pi \bm\sigma\hat\pi$ in (\ref{kakav}), and deduce the time-invariance of $\rho_s$. 

Thus, we have two non-relativistic forms of energy, $\rho$ and $\rho_s$ that are conserved locally and expressed via the Schroedinger wave function $\phi$ from (\ref{er}). Note from (\ref{den}, \ref{gru}, \ref{len}, \ref{sta}) that $mc^2(n_0-\phi^\dagger\phi)={\cal O}(1)$ is also locally conserved, 
and has the non-relativistic order of magnitude. But it is not expressed only via $\phi$, i.e. it demands solving a quasi-relativistic equation for $\varphi$; see (\ref{gro}). Hence only (\ref{gru}) and (\ref{kakav}) define new observables for non-relativistic quantum mechanics. We focus on $\rho$ and $\rho_s$, which we regard as different forms of energy. 

\section{Stationary states with non-zero holographic energy (\ref{kakav})} 
\label{stato}

Recall that $\rho_s=0$ whenever the stationary $\phi(\bm r)$ is real. Likewise, $\rho_s=0$ if $\phi(\bm r)$ is a plane wave. Hence we focus on stationary scattering states that do describe an infinite motion, and we need to look for interference effects there that go beyond a single plane wave. The spin is crucial for the existence of (\ref{kakav}), but it is sufficient to assume the simplest factorized situation $\phi_s(\bm r)=|s\rangle \phi(\bm r)$. Thus, (\ref{kakav}) will contain the mean magnetization $\bm \mu\equiv\bra{s}\bm \sigma\ket{s}$. Now for $\phi(\bm r)$ we assume the assume the following stationary solution $\phi(\bm r,t)=e^{-itE/\hbar} \phi(\bm r)$ of (\ref{er}):
\BEA
\label{berg}
\phi(\bm r)=e^{ikz}+\frac{fe^{ikr}}{r}, ~~ r=|\bm r|,~~E=\frac{\hbar^2k^2}{2m},
\EEA
where $\bm r=(x,y,z)$, $E>0$ is the energy, and $k$ is the wave-vector. Eq.~(\ref{berg}) describes the incident wave $e^{ikz}$ that is scattered on a potential $U(\bm r)\propto \delta(\bm r)$ centered at $\bm r= 0$ \cite{landau,davydov}. This produced the scattered expanding spherical wave in (\ref{berg}), where $f$ is a constant scattering amplitude. We get from (\ref{kakav}, \ref{berg}):
\begin{equation}\begin{split}
\rho_s&=\frac{\hbar^2k^2(\mu_xy-\mu_yx)f}{2mr^2}\times\\
&\times\Big[\sin(kz-kr)-\frac{\cos(kz-kr)}{kr}
\Big],\\
\bm \mu&\equiv\bra{s}\bm \sigma\ket{s}. 
\label{vovk}
\end{split}\end{equation}
Note that the interference between the two waves in (\ref{berg}) is essential for $\rho_s\not=0$ in (\ref{vovk}). We see in (\ref{vovk}) that ${\cal R}^2\oint\d \Omega\, \rho_s=0$, where $\int\d\Omega$ is the surface integral over the sphere with radius ${\cal R}$. However, if we integrate over a part of this sphere, ${\cal R}^2\int\d \Omega\, \rho_s$ can scale as ${\cal O}({\cal R})$. In that sense $\rho_s$ concentrates on the spherical surface. 

For a non-zero magnetic field $\bm B\not=0$ the stationary wave-function $\phi$ need not be real. (Note that in the presence of a stationary magnetic field all the space-derivatives in the energy density are changed to gauge-invariant derivatives defined via the vector potential). Hence $\rho_s\not=0$ in (\ref{kakav}) is possible for normalizable stationary states. An important example of this type is provided by Landau levels for a weakly-confined 2d electron gas under constant and homogeneous $\bm B$; see \cite{tong} for review. This system is basic e.g. for the quantum Hall effects and related macroscopic quantum states \cite{tong}. Appendix \ref{aa5} studies this system, shows that $\rho_s\not=0$ due to $\bm B\not=0$, and that $\rho_s$ concentrates at the surface of the system. 

\section{Summary} 

The concepts of energy density and energy current were addressed in literature many times. The definitions of these quantities are not unique due to non-commutativity. Considering the diversity of opinions, there are two ways to develop these concepts: show which definition emerges from more fundamental physics, and produce results independent of definitional differences. Both these approaches were followed in this paper. 

We show that the energy density and current in quantum mechanics can be defined consistently with fundamental relativistic physics and that this definition coincides with that provided by Terletsky-Margenau-Hill coordinate-momentum quasiprobability. We applied this result to deducing the energy transfer velocity for two classes of localized wave packets (including Gaussian states) and showing that it generically exceeds the coordinate transfer (i.e. group) velocity. The structure of this result depends only on the space-integrated energy current. Hence it does not depend on how precisely one defines the energy density.

Further interesting questions stay open; e.g. the energy motion in tunneling or the extension of our results to open, discrete systems, as was recently done for the probability current \cite{seneca}. We also uncovered a new form of energy with a non-relativistic magnitude which is essentially spin-dependent. It is holographic, i.e. it does not contribute to the global energy budget of finite-motion states, but its local contribution is sizable. We illustrated this spin-dependent energy for two physically pertinent examples: stationary scattering and Landau levels of electrons in a magnetic field.

The concepts of energy density and current are important in condensed matter physics, where they define, in particular, heat currents \cite{mclennan,hardy,canadian,muga05,Wu_2009,juan10,sanchez14,tachibana01,geraldine17}. Also in this field, different definitions are employed; cf. e.g. Refs.~\cite{canadian,hardy,mclennan}. Hence, there is room for advancing these subjects; e.g. in the future we plan to reconsider the heat and energy current for phonon systems within the formalism of Ref.~\cite{we}.

\begin{acknowledgments}
This work was supported by SCS of Armenia, grants 20TTAT-QTa003, 21AG-1C038 and 22AA-1C023. We thank K. Hovhannisyan for discussions. 
\end{acknowledgments}

\bibliographystyle{quantum}
\bibliography{main}

\appendix

\section{Tranformation of energy density under Galilean boost}
\label{boost-a}
{\color{black}
Define the Galilean boost as 
\BEA
\label{boost}
&&\bm r=\bm r'+\bm V t,\qquad t=t',\\
&&\bm v=\bm v'+\bm V,
\label{oost}
\EEA
where $\bm V$ is the relative velocity of two reference frames ${\cal S}$ and ${\cal S}'$. 
It is well known that the Schroedinger wave function transforms under (\ref{boost}) via a phase factor:
\BEA
\phi(\bm r, t)=\phi'(\bm r', t)\exp\Big[ \frac{im}{\hbar}\Big( \bm V\bm r-\frac{t}{2}\bm V^2 \Big)   \Big ].
\label{boost2}
\EEA
Eq.~(\ref{boost2}) can be derived directly from the Schroedinger equation, or alternatively, via the plane-wave expansion of a given solution.
We get from (\ref{oliki}, \ref{boost}, \ref{boost2})
\BEA
\label{rashida}
&&\rho(\bm r,t)=
\rho'(\bm r',t)+\frac{m\bm V^2}{2}\phi'{}^{\dagger}(\bm r')\phi'(\bm r') \\
&&+\bm V\,\Re[\phi'{}^{\dagger}(\bm r') \frac{\hbar}{i}\bm\nabla_{{\bm r}'}\phi'(\bm r')],
\label{vitte} 
\EEA
where in (\ref{vitte}), $\bm V$ is multiplied by the momentum density, as verified via the local conservation law of the coordinate density.
Eqs.~(\ref{rashida}, \ref{vitte}) are the local generalizations of 
\BEA
\frac{m\bm v^2}{2}=\frac{m\bm v'^2}{2}+\frac{m\bm V^2}{2}+m\bm v'\bm V,
\EEA
which follows from (\ref{oost}). Interestingly, $\widetilde\rho$ from (\ref{boliki}) holds the same transformation (\ref{rashida}, \ref{vitte}) with $\rho$ replaced by $\widetilde\rho$. This confirms that both $\rho$ and $\widetilde\rho$ are consistent with the same momentum density defined via the Terletsky-Margenau-Hill quasi-probability for the coordinate and momentum, i.e. in a sense there is no controversy on what is the momentum density in non-relativistic quantum mechanics, though momentum and coordinate do not commute. 
}

\section{Derivation of Dirac's energy density and current from the relativistic Lagrangian and the energy-momentum tensor }
\label{aa1}

The relativistic Lagrangian of spin-$1/2$ particle reads ($\hbar=c=1$) \cite{bere}:
\BEA
\label{1}
{\cal L}=\frac{i}{2}\left(\bar{\psi}\gamma^\mu\partial_\mu\psi-\partial_\mu\bar{\psi}\gamma^\mu\psi\right)-m\bar{\psi}\psi,
\EEA
where $m$ is the mass, $\psi$ is the four-component column bispinor, $\mu=0,...,4$, $A_\mu B^\mu=A^0B^0-\bm A\cdot\bm B$, $\partial_\mu=(\partial_t,\vec{\nabla})$, and $\gamma^\mu$ are gamma matrices [see (\ref{fantom})]: 
\begin{equation}
\gamma^0=\beta,\qquad \gamma^k=\beta\alpha_k,\qquad i,j,k=1,2,3,
\end{equation}
\begin{equation}
\alpha_i\alpha_j+\alpha_j\alpha_i=2\delta_{ij}, \quad \alpha_k\beta+\beta\alpha_k=0,\quad
\beta^2=1.
\label{go}
\end{equation}
Recall that (\ref{1}) is a 4-scalar \cite{bere}. 
Within Dirac's representation $\vec{\alpha}$ and $\beta$ are given by (\ref{fantom}).

The energy-momentum tensor reads from (\ref{1})
\BEA
\label{2}
T^{\mu}_{\quad \nu}=\frac{\partial {\cal L}}{\partial [\partial_\mu\psi]}\partial_\nu\psi
+\partial_\nu\bar{\psi} \frac{\partial {\cal L}}{\partial [\partial_\mu\bar{\psi}]}-\delta^{\mu}_{\nu}{\cal L}\nonumber\\
=\frac{i}{2}\left(\bar{\psi}\gamma^\mu\partial_\nu\psi-\partial_\nu\bar{\psi}\gamma^\mu\psi\right).
\EEA
Note that $T^{\mu}_{\quad \nu}\not=T^{\nu}_{\quad \mu}$. It holds four conservation laws:
\BEA
\label{3}
\partial_\mu T^{\mu}_{\quad \nu}=0.
\EEA
We shall look at the energy conservation. The energy density and energy current reads from (\ref{2}):
\begin{equation}
T^{0}_{\quad 0}=-\frac{i}{2}\left({\psi}^\dagger {\alpha_k}{\nabla_k}\psi-{\nabla_k}{\psi}^\dagger{\alpha_k}\psi\right)
+m\psi^\dagger\beta\psi,
\end{equation}
\begin{equation}
T^{i}_{\quad 0}=-\frac{i}{2}\left({\psi}^\dagger {\alpha}_i\alpha_j\partial_j\psi
-\partial_j{\psi}^\dagger\alpha_j{\alpha}_i\psi\right)
\label{8}
\end{equation}
where $i=1,2,3$. Using in (\ref{8}):
\begin{equation}\begin{split}
& {\psi}^\dagger {\alpha}_i\alpha_j\partial_j\psi
-\partial_j{\psi}^\dagger\alpha_j{\alpha}_i\psi = {\psi}^\dagger \partial_i\psi\\
& 
-\partial_i{\psi}^\dagger\psi +\sum_{j\not=i}\left( {\psi}^\dagger {\alpha}_i\alpha_j\partial_j\psi
-\partial_j{\psi}^\dagger\alpha_j{\alpha}_i\psi\right),
\end{split}\end{equation}
and noting from (\ref{go}) 
\BEA
{\alpha}_i\alpha_j=i\epsilon_{ijk}\tau_k, \qquad 
{\tau_i}=
\begin{pmatrix}
{\sigma_i} & 0 \\
0 & {\sigma_i} 
\end{pmatrix},
\EEA
we see that the 3-vector of energy current in (\ref{8}) is a sum of two contributions:
\BEA
\label{12}
T^{i}_{\quad 0}=\bm{{\cal J}}+\frac{1}{2}{\bm \nabla}\times(\psi^\dagger{\bm \tau}\psi),\\
\bm{{\cal J}}=-\frac{i}{2}(\psi^\dagger{\bm \nabla}\psi  -{\bm \nabla}\psi^\dagger\psi   ),
\label{13}
\EEA
where $\times$ means vector product. The contribution $\frac{1}{2}{\bm\nabla}\times(\psi^\dagger{\bm \tau}\psi)$ in (\ref{12}) is interesting, but we shall not focus on it, since it is a rotor. Finally, we can write the energy conservation in the dimensional form, also incuding there the potential energy $U$; see (\ref{vava}, \ref{vovo}). 

\section{Lagrangian perspective of the non-relativistic energy density and current }
\label{aa2}

The energy density \eqref{gru} for the Schroedinger wave-function $\phi$ [cf.~(\ref{er})] can be found via the following Lagrangian density
    \begin{equation}\label{eq:SchrodingerTransformedLagrangian}
        \mathcal{L} = \frac{\hbar^2}{4m} \bigg(\phi^\dagger\Delta\phi + \Delta\phi^\dagger\phi\bigg) - U\phi^\dagger\phi + \frac{i\hbar}{2}(\phi^\dagger\dot\phi - \dot\phi^\dagger\phi).
    \end{equation}
For simplicity we assume that the potential $U(\bm r)$ is time-independent. 
Now $\mathcal{L}$ contains only second-order space-derivatives. Equations of motion generated 
by $\mathcal{L}$, 
\BEA
\frac{\d}{\d t}\frac{\partial\mathcal{L} }{\partial \dot q}=
\frac{\partial\mathcal{L}}{\partial q}+\sum_{k=1}^3\partial^2_{kk}
\frac{\partial\mathcal{L}}{\partial [\partial^2_{kk}{q}]}, \qquad
q=\phi, \phi^\dagger, 
\label{lagr}
\EEA
coincide with Schrodinger's equation (\ref{er}). Below we shall omit the summation sign $\sum_{k=1}^3$.

Using (\ref{lagr}) we verify that the local energy conservation law  generated due to $\partial_t\mathcal{L}=0$,
\BEA
&&\frac{\d}{\d t} \Big[
 \frac{\partial\mathcal{L} }{\partial \dot \phi} \dot \phi
+\dot \phi^\dagger \frac{\partial\mathcal{L} }{\partial \dot \phi^\dagger}
-\mathcal{L}\Big]\nonumber\\ 
&&+\partial_k\Big[
\frac{\partial\mathcal{L}}{\partial [\partial^2_{kk}{\phi}]}\partial_k\dot{\phi}
-\partial_k\Big( \frac{\partial\mathcal{L}}{\partial [\partial^2_{kk}{\phi}]}\Big) \dot \phi
\nonumber\\
&&+\Big(\partial_k\dot{\phi}^\dagger\Big)\frac{\partial\mathcal{L}}{\partial [\partial^2_{kk}{\phi^\dagger}]}
-\dot \phi^\dagger\partial_k \frac{\partial\mathcal{L}}{\partial [\partial^2_{kk}{\phi^\dagger}]}
\Big]=0,
\label{lagr2}
\EEA
coincides with (\ref{gru}, \ref{bonbon}). More precisely for current generated from (\ref{lagr2}) we get
[cf.~(\ref{bonbon})]
\BEA
{\bm J}=\frac{\hbar^2}{2m}\Re\Big[ \phi^\dagger\bm\nabla\dot \phi -[\bm\nabla\phi^\dagger]\dot\phi    \Big],
\label{bonbon2}
\EEA
where one can employ the Schroedinger equation (\ref{er}). Note that the non-relativistic limit taken in (\ref{13}) produces a different current:
\BEA
\label{bonbon3}
&\bm J_{\rm D} = \frac{i\hbar U}{2m}(\bm\nabla\phi^\dagger\phi-\phi^\dagger\bm\nabla\phi) + \frac{i\hbar^3}{8m^2}(\phi^\dagger\bm\nabla\Delta\phi \nonumber\\
       &- [\partial_i\phi^\dagger]\bm \nabla\partial_i\phi + \bm \nabla\partial_i\phi^\dagger[\partial_i\phi] - [\bm\nabla\Delta\phi^\dagger]\phi).
\EEA
However, the difference between (\ref{bonbon2}) and (\ref{bonbon3}) amounts to a rotor as verified directly. I.e. this is an allowed difference. Recall that one rotor was already neglected in (\ref{13}) compared with (\ref{12}).

Note that the choice of the non-relativistic Lagrangian density
(\ref{eq:SchrodingerTransformedLagrangian}) is not unique. The usual,
first-order space-derivative containing density is \cite{greiner}
    \begin{equation}\label{budker}
        \Lambda= -\frac{\hbar^2}{2m} {\bm\nabla}\phi^\dagger{\bm\nabla}\phi -
        U\phi^\dagger\phi + \frac{i\hbar}{2}(\phi^\dagger\dot\phi - \dot\phi^\dagger\phi).
    \end{equation}
Eq.~(\ref{budker}) generates the same (\ref{er}):
\BEA
\frac{\d}{\d t}\frac{\partial\Lambda }{\partial \dot q}=
\frac{\partial\Lambda }{\partial q}-\partial_{k}
\frac{\partial\Lambda }{\partial [\partial_{k}{q}]}, \qquad
q=\phi, \phi^\dagger.
\label{lagr7}
\EEA

The difference between (\ref{budker}) and (\ref{eq:SchrodingerTransformedLagrangian}) is that the former contains only first space-derivatives, while the latter only second-order space-derivatives. The very possibility of such a difference is due to non-relativism, where the space and time enters differently. Recall in this context that the relativistic Dirac's Lagrangian (\ref{1}) which we write again as
\BES
        \mathcal{L} &= \frac{ihc}{2}(\psi^\dagger\bm\alpha \bm \nabla\psi-\bm \nabla\psi^\dagger\bm \alpha\psi) + \frac{i\hbar}{2}(\psi^\dagger\dot\psi-\dot\psi^\dagger\psi) \\
        &-mc^2\psi^\dagger\beta\psi - U\psi^\dagger\psi,
        \label{kushan}
\EES 
does not contain such a non-uniqueness, since the time and space enters into (\ref{kushan}) equivalently, i.e. via the first derivatives over space and time \cite{bere}.

Returning to $\mathcal{L}$ and $\Lambda$, we note that 
\BEA
\int\d^3 x\, \mathcal{L}=\int\d^3 x\,\Lambda.
\EEA
Now $\int\d^3 x\,\Lambda$ equals the Dirac-Frenkel Lagrangian \cite{frenkel}. We do not suggest defining the stationary action principe via $\mathcal{L}$ or $\Lambda$. Such a principle is problematic (and needs additional considerations) for the first order (in time) Schroedinger equation \cite{leeuwen,vignale}. For us $\mathcal{L}$ or $\Lambda$ are tools for studying the symmetries, including time-translation invariance
and the ensuing local energy conservation. 

Eq.~(\ref{lagr7}) leads to the local energy conservation provided that $\Lambda$ does not depend on time explicitly:
\BEA
\label{yezidi}
\partial_t\Big[ \frac{\partial\Lambda}{\partial \dot{q}} -L \Big]+\partial_k
\Big[ \dot q \frac{\partial\Lambda}{\partial [\partial_kq]}\Big]=0,
\EEA
which reads for (\ref{budker}):
\begin{align}
&\frac{\d}{\d t} \Big[
 \frac{\partial\Lambda }{\partial \dot \phi} \dot \phi
+\dot \phi^\dagger \frac{\partial\Lambda }{\partial \dot \phi^\dagger}
-\Lambda\Big]\nonumber\\ 
&+\partial_k\Big[
\dot{\phi}^\dagger\frac{\partial\Lambda}{\partial [\partial_{k}{\phi}^\dagger]}
+\Big( \frac{\partial\Lambda}{\partial [\partial_{k}{\phi}]}\Big) \dot \phi
\Big]=0, \label{lagr3} \\
\label{porto}
&\widetilde{\rho}=\frac{\hbar^2}{2m} {\bm\nabla}\phi^\dagger{\bm\nabla}\phi +U\phi^\dagger\phi\\
& =\frac{1}{2m}\bra{\bm{r}}\bm{P}R\bm{P}\ket{\bm{r}}
+ \bra{\bm{r}}R U\ket{\bm{r}}, \quad \bm{P}\equiv-i\hbar\bm\nabla.
\label{dag}
\end{align}
This is a different local energy density.
This candidate for energy density has been discussed recently in \cite{Francisco23}, where the authors study an infinite square potential well with moving walls. They argue that the positive energy density \eqref{porto} correctly predicts the average work done by the motion of the walls while the energy density \eqref{oliki} does not make this prediction because it nullifies at the wall. While this assessment is mathematically correct, we believe it came about due to the singular nature of the considered potential. For smooth potentials, \eqref{oliki} correctly predicts the local supply of work. This is seen from the continuity equation \eqref{ord}, which contains a work term $\dot U \phi^\dagger\phi$. 
In particular, if the energy current \eqref{bonbon3} is zero we will have
\begin{equation}
    \dot \rho (\bm{r}, t) = \dot U(\bm{r}, t) \phi^\dagger\phi.
\end{equation}
This is a correct prediction for local work not yet redistributed by a current of energy.

\section{Gaussian wavefunctions }
\label{aa3}

The Schroedinger equation for a single-particle wavefunction evolving under the free propagation Hamiltonian $\hat{H} =\frac{1}{2 m} \hat{p}^2$ in 1d is easily solved in the momentum representation: 
\BEA
\label{ford}
i\hbar\partial_t \phi(p,t)=\frac{p^2}{2m}\phi(p,t). 
\EEA
Eq.~(\ref{ford}) leads to
\begin{equation}
\phi(p,t)=e^{-itp^2/(2m\hbar)}\chi(p), \label{eq:psipt}
\end{equation}
where $\chi(p)$ is a normalized function, $\int \d p |\chi(p)|^2 = 1$. 
For Gaussian wavefunctions we take $\chi(p)$ to be
\begin{equation} 
\label{eq:chip}
\chi(p) = \frac{\sqrt{a}}{\pi^{\frac{1}{4}}} \exp \left[-\frac{a^2}{2}(p-\frac{b}{a})^2\right],~~
\int \d p\, |\chi(p)|^2=1,~
\end{equation}
with constants $a>0$ and $b$ defined via the mean and variance of the momentum:
\begin{equation}
\braket{p}\equiv\int\d p\,p|\chi(p)|^2=\frac{b}{a}, \qquad
\braket{p^2}-\braket{p}^2=\frac{1}{2a^2}.
\label{gustaf}
\end{equation}
Eq.~ (\ref{eq:chip}) defines a general pure Gaussian state. There can be two more real parameters $\alpha$ and $\beta$ in (\ref{eq:chip}) in the sense that it can be multiplied by $e^{i\alpha p+i\beta p^2}$. However, $\alpha$ and $\beta$ shift (respectively) the initial coordinate and initial time that for (\ref{eq:chip}) are both taken to be zero. 

    \begin{figure}[h!]
        \includegraphics[clip,width=1\linewidth]{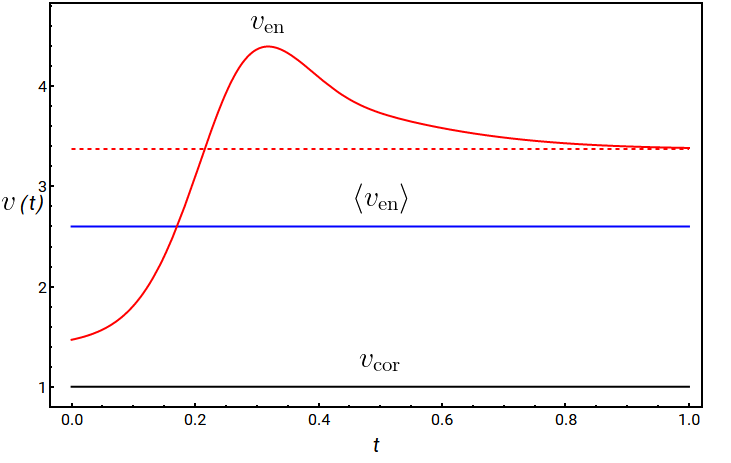}
        \caption{Energy transfer velocities $v_{\rm en}$ and $v_{\rm mp}$, and the coordinate transfer velocity $v_{\rm cor}$ for the Gaussian wave-packet (\ref{eq:psixt}); see (\ref{du1}, \ref{du2}, \ref{du3}). Parameters assume the following values: $a = 2^{-1/2}$, $m = 2^{-1}$, $b = 2^{-3/2}$, $\hbar = 1$.}
        \label{fig_gauss}
    \end{figure}
    
    \begin{figure}[h!]
        \includegraphics[clip,width=1\linewidth]{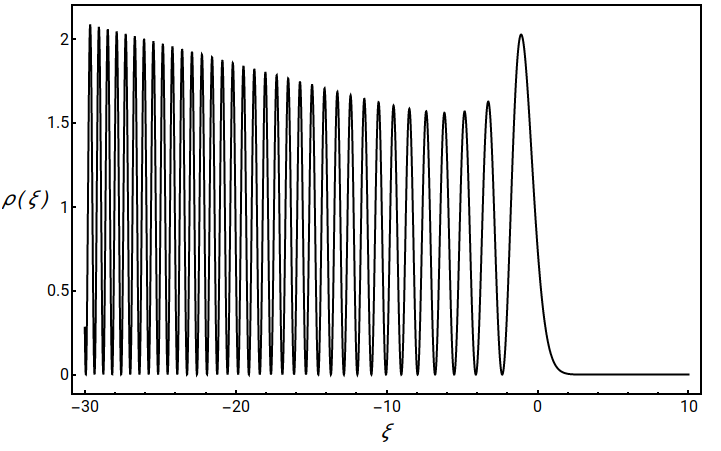}
        \caption{For the Airy wave-function (\ref{airywavefunction}) we show its energy density (\ref{airyenergydensity})
        as a function of $\xi=\xi_t$. Parameters are given as: $\beta=\hbar=m=1$, $t = \sqrt{6}$. We see that the energy density has a pronounced (first from the right) local maximum near the wave-front at $\xi=0$. However, there is no a finite most probable value of energy, since it grows indefinitely for $\xi\to-\infty$.}
        \label{fig:energy_density_airy}
    \end{figure}

\begin{figure}[h!]
        \includegraphics[clip,width=1\linewidth]{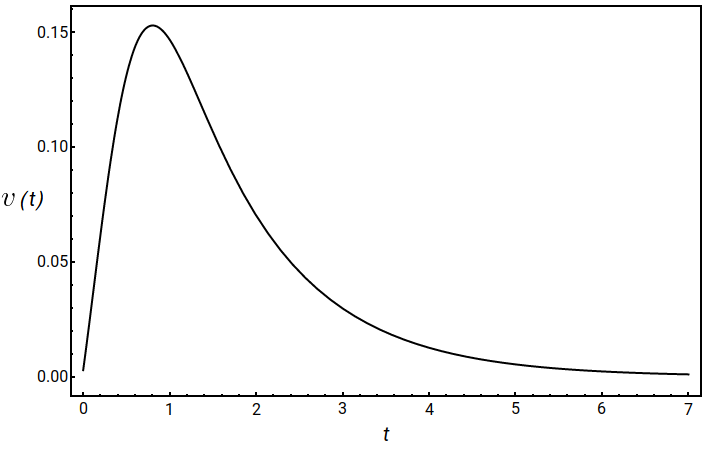}
        \caption{The difference $v_r(t)=v_{\rm en}(t)-v_{\rm cor}(t)$ between the energy transfer velocity and coordinate transfer velocity [see (\ref{airycorgroupvel}, \ref{buka})] for the Airy wave-function (\ref{airywavefunction}). 
        Parameters read: $\beta=\hbar=m=1$. }
        \label{fig:airy_energy_group_velocity}
    \end{figure}

Recalling the $\delta$-function normalized momentum eigenfunctions, 
$\braket{x | p} = \frac{1}{\sqrt{2 \hbar\pi}}\, e^{i p x/\hbar}$, we obtain 
from Eq.~\eqref{eq:chip} the wavefunction in the coordinate representation:
\begin{equation}\begin{split}
&\phi(x, t) = \frac{1}{\sqrt{2 \pi\hbar}} \int \d p \, e^{ipx/\hbar}\phi(p,t)\\
\label{eq:psixt}
&= \frac{1}{\pi^{\frac{1}{4}}\sqrt{a (1 + i \tau)}} \exp\left[-\frac{b^2}{2} - \frac{(\xi - i b)^2}{2(1 + i \tau)} \right]\\
&= \frac{1}{\pi^{\frac{1}{4}}\sqrt{a (1 + i \tau)}}\times\\&\times \exp\left[-\frac{(\xi-b\tau)^2}{2(1+\tau^2)} + \frac{i}{2}\,\,\frac{2b\xi+\tau(\xi^2-b^2)}{1+\tau^2} \right], 
\end{split}\end{equation} 
where 
\BEA
\xi \equiv x/(a\hbar), \qquad \tau \equiv t/(ma^2\hbar), 
\EEA
are the dimensionless coordinate and time, respectively.

For the energy density, particle current, and energy current we get, respectively,
\begin{align}
&\rho=-\frac{1}{2ma^2}\Re[\phi \frac{\d^2 \phi^*}{\d \xi^2}]\\
\label{rable}
&=\frac{|\phi|^2}{2ma^2}\, \frac{\left(\tau^2-1\right) \left(\xi ^2-b^2\right)+4 b \tau \xi +\tau^2+1}{\left(\tau^2+1\right)^2},\\
\label{fr}
&|\phi|^2 v=\frac{1}{ma}\Im[\phi^* \frac{\d \phi}{\d \xi}],\\
\label{franc}
&v=\frac{1}{ma}\,\frac{b+\xi\tau}{1+\tau^2},\\
&J=-\frac{1}{4m^2a^3}\Im[\phi^* \frac{\d^3 \phi}{\d \xi^3}+\frac{\d^2 \phi^*}{\d \xi^2}\frac{\d \phi}{\d \xi}     ]  \\
&= \frac{|\phi|^2}{4m^2a^3(1+\tau^2)^3}\Big[2\tau(\tau^2-1)\xi^3+2b(5\tau^2-1)\xi^2\nonumber\\
&+2\tau(3(1+\tau^2)+b^2(\tau^2-5))\xi\nonumber\\
&+2b(2+(1-\tau^2)(b^2+\tau^2) \Big].
\end{align}
Note from (\ref{rable}) that for $b=0$, $\rho(\xi,\tau)>0$ for all $\xi$ and $|\tau|>1$; i.e. $\rho(\xi,\tau)<0$ only for a finite interval of $\tau$. Thus (\ref{rable}) provides an example, when the energy density $\rho(\bm r,t)$ is negative only for a limited range of $\bm r$ and $t$. In the main text we conjectured that such a limited negativity of $\rho(\bm r,t)$ is unavoidable for any normalizable state of the free quantum motion; see section \ref{negus}. 

Eqs.~(\ref{fr}, \ref{franc}) show that the coordinate transfer velocity can be defined either via the mean current $\int\d\xi |\phi|^2 v$ or via $v$ calculated at the maximally probable value $\xi=b\tau$ of the Gaussian wave-packet. In both cases the coordinate transfer velocity reads [cf.~(\ref{gustaf})]
\BEA
\label{du1}
v_{\rm cor}=\frac{b}{ma}=\frac{\braket{p}}{m}.
\EEA
The definition (\ref{prangle}) of the energy transfer velocity $v_{\rm en}$ 
is worked out directly from (\ref{gustaf}): 
\BEA
\label{du2}
v_{\rm en}-v_{\rm cor}=\frac{b}{ma}\frac{1}{1+2b^2}\geq 0.
\EEA
It is seen that $v_{\rm en}-v_{\rm cor}\to 0$ both for low-speed and high-speed situations, i.e. for $b\to 0$ and $b\to\infty$. 

Another pertinent definition of energy transfer velocity, $v_{\rm mp}(t)$, is given by looking at the maximum of the energy density (\ref{rable}) over $\xi$ for fixed $\tau$ (i.e. over $x$ for fixed $t$). Denote this maximum as $\xi_{\rm mp}(\tau)$ and then define:
\BEA
\label{du3}
v_{\rm mp}=\frac{1}{ma}\,\frac{\d\xi_{\rm mp}}{\d\tau}.
\EEA
Fig.~\ref{fig_gauss} compares with each other $v_{\rm mp}(t)$ (which is time-dependent), $v_{\rm en}$, and $v_{\rm cor}$. As expected, $v_{\rm en}>v_{\rm cor}$ and $v_{\rm mp}(t)>v_{\rm cor}$, but the relation between $v_{\rm en}$ and $v_{\rm mp}(t)$ is more intricate. 

\section{Airy wave packet}
\label{aa4}

The free Airy wave-packet is given by a non-normalizable wave-function that has two interesting properties \cite{berry}. First, it moves without changing its form, i.e. it moves without dispersion. Second, it moves with a constant acceleration, although it is not subject to any force. These features have to do with its non-normalizable character, but (as expected) both will survive at least for some times for suitably normalizable analogs of the free Airy wave-packet that were observed experimentally \cite{observation}.

The Airy wave function \cite{berry} solves the free Schroedinger equation (\ref{er}):
\begin{equation}
\label{airywavefunction}
\phi(x,t) = \Ai\bigg(\frac{\beta}{\hbar^{2/3}}\,\xi_t\bigg)\exp\bigg[\frac{i}{\hbar}t\,\frac{\beta^3}{2m}\bigg(\xi_t + \frac{\beta^3t^2}{12m^2}\bigg)\bigg],
\end{equation}
\begin{equation}
\xi_t \equiv x - \frac{\beta^3t^2}{4m^2},
\label{nepal}
\end{equation}
where $\beta>0$ is a constant, and 
$\Ai(x)$ is the Airy function. It is a particular solution of the linear differential equation with two conditions: 
\begin{equation}
\label{airo}
\frac{\d^2}{\d x^2}\Ai\equiv \Ai''[x]=x \Ai[x], ~~~ \Ai[|x|\to\infty]\to 0.
\end{equation}
For $x\gg 1$, $\Ai[x]$ goes to zero quickly, while its decay is very slow (and hence $\Ai[x]$ is not normalizable) for $x\ll- 1$:
\begin{equation}
\label{vi}
\Ai[x\ll -1]\simeq \frac{1}{\sqrt{\pi}\,(-x)^{1/4}}\sin\Big[\frac{\pi}{4}+\frac{2}{3}(-x)^{3/2}\Big],
\end{equation}
\begin{equation}
\Ai[x\gg 1]\simeq \frac{1}{\sqrt{\pi}\,x^{1/4}}\exp\Big[-\frac{2}{3}x^{3/2}\Big],
\label{gust}
\end{equation}
Note that (\ref{airo}) can be visualized as the Newton equation of motion for a particle subject to a time-dependent harmonic potential $-\frac{tX^2}{2}$, where $t\to x$ and $X\to \Ai$ for (\ref{airo}). This potential changes from confining (at $t<0$) to deconfining for $t>0$. In this physical sense the oscillatory behavior (\ref{vi}) is more expected than (\ref{gust}).

Thus the density $|\phi(x,t)|^2$ of the wave-packet (\ref{airywavefunction}) has an oscillatory tail extending to $\xi_t\to-\infty$ and a sharp wave-front around $\xi_t\simeq 0$. The global maximum of $|\phi(x,t)|^2$ is located around this wave-front that moves with acceleration according to (\ref{nepal}). 

The energy density of the Airy wave-packet (\ref{airywavefunction}) is deduced from (\ref{gru}, \ref{airo}):
        \begin{equation}
        \label{airyenergydensity}
            \rho(x,t) = \frac{\beta^3}{2m}\bigg[\frac{\beta^3t^2}{4m^2} - \xi_t\bigg]\Ai^2\bigg(\frac{\beta}{\hbar^{2/3}}\xi_t\bigg).
        \end{equation}
It is clear from (\ref{airyenergydensity}, \ref{vi}) that the energy density grows (with oscillations) for $x\to-\infty$; see Fig.~\ref{fig:energy_density_airy}. This is the main qualitative difference between $\rho(x,t)$ and $|\phi(x,t)|^2$. Such an infinite energy reservoir is the physical reason for the acceleration behavior of the Airy wave-packet. 

We turn to the coordinate group velocity of \eqref{airywavefunction}. We can define this as the velocity $v_{\rm cor}(t)$ of the most probable value of $|\phi(x,t)|^2$, which is located near the wave-front. 
Since $|\phi(x,t)|^2$ depends only on $\xi_t$, we find
        \begin{equation}
        \label{airycorgroupvel}
            v_{\rm cor}(t) =-\dot\xi_t= \frac{\beta^3}{2m^2}t,
        \end{equation}
which again confirms that \eqref{airywavefunction} moves with a constant acceleration. Since $|\phi(x,t)|^2$ for the Airy wave-packet moves as a whole without dispersion, $v_{\rm cor}(t)$ can be identified via the velocity of any of its points. Note the following relation between (\ref{airycorgroupvel}) and (\ref{airyenergydensity}):
        \begin{equation}
            \partial_t{\rho}(y, t)|_{y=\xi_t} = |\phi(x,t)|^2\frac{\d}{\d t}\frac{mv_{\rm cor}^2(t)}{2}.
        \end{equation}
The energy transfer velocity for the Airy wave-packet can be defined as the velocity of the local maximum of $\rho(x,t)$ [given by (\ref{airyenergydensity})]. One first maximizes $\rho(x,t)$ as a function of $x$ for a given $t$, denotes this by $x_{\rm en}(t)$ and then defines the energy transfer velocity as:
\BEA
\label{buka}
v_{\rm en}(t)=\dot x_{\rm en}(t). 
\EEA
Comparing (\ref{buka}) with (\ref{airycorgroupvel}) we see that $v_{\rm en}(t)\geq v_{\rm cor}(t)$ with         
$v_{\rm en}(t)\to v_{\rm cor}(t)$ for long times; see Fig.~\ref{fig:airy_energy_group_velocity}.  

\section{An example where the energy is transferred slower than the coordinate}
\label{simon}

Returning to the content of section \ref{macro-energy-transfer}, consider the following momentum probability density, which is tailed towards $P<0$:
\begin{align}
\label{ben}
f(P)&= a\,e^{a(P-A)}~~{\rm for}~~P\leq A,\\
&=0~~{\rm for}~~P> A,
\label{mordokhai}
\end{align}
where $a>0$ and $A>0$ are parameters. 
Eqs.~(\ref{ben}, \ref{mordokhai}) imply
\begin{align}
& \braket{P}=\int\d P\, Pf(P)= \frac{aA-1}{a} \\
& \braket{P^2}= \frac{(aA-1)^2+1}{a^2} \\
& \braket{P^3}= \frac{a A (a A (a A-3)+6)-6}{a^3}, \\
& \braket{P^3}-\braket{P^2}\braket{P}=    \frac{2 (a A-2)}{a^3},
\end{align}
where the last expression defines the sign of the difference between the energy transfer velocity and the coordinate transfer velocity; see (\ref{prangle}). Now for $2>aA>1.5961$ we get $\braket{P}>0$ and $\braket{P^3}>0$ -- i.e. the energy amd coordinate flow in the same direction, as demanded by (\ref{komma}) -- but $\braket{P^3}-\braket{P^2}\braket{P}<0$, i.e. the energy flow is slower.

\section{The rest energy and negativity of the non-relativistic energy}
\label{resto}

We saw above that the kinetic part of $\rho(\bm r, t)$ can be negative. Moreover, we conjectured that some negativity is inevitable for any smooth and normalizable pure state for $t>0$ or for $t<0$ (or for both); see section \ref{negus}. Here we discuss (admittedly rather formally) whether this negativity is guaranteed to leave positive the full energy density $\varrho(\bm r, t)$, which in the non-relativistic limit is given by (\ref{eq:DiracDensityCalcd}). First, we rewrite (\ref{eq:DiracDensityCalcd}) as
\begin{align}
\label{gaza1}
\varrho =& \rho+ mc^2\chi^\dagger\chi +mc^2(\phi^\dagger\kappa +\kappa^\dagger\phi) \\ 
+& mc^2\phi^\dagger \phi + \mathcal{O}(c^{-2}),
\label{gaza2}
\end{align}
where $\rho$ is given by (\ref{gru}), and $\kappa\equiv \varphi-\phi={\cal O}(c^{-2})$, as seen from (\ref{susanna}). In (\ref{gaza1}, \ref{gaza2}) all factors besides $\rho$ are contributions into the rest energy. We note from (\ref{susanna}, \ref{gla}) that all factors in (\ref{gaza1}) are or order of ${\cal O}(1)$, while the rest energy density in (\ref{gaza2}) is ${\cal O}(c^2)$. Let us now compare the kinetic energy density $-\frac{\hbar^2}{2m}\Re[\phi^\dagger\Delta \phi]$ with $mc^2\phi^\dagger\phi$. Using the Cauchy-Schwartz inequality (\ref{ober}) we find (\ref{maj}):
\begin{equation}
\label{ober}
|\Re[\phi^\dagger\Delta \phi]|\leq |\phi^\dagger\Delta \phi|\leq \sqrt{\phi^\dagger \phi}\,\sqrt{\Delta\phi^\dagger\Delta\phi},
\end{equation}
\begin{equation}
\label{maj}
mc^2\phi^\dagger\phi-\frac{\hbar^2}{2m}|\Re[\phi^\dagger\Delta \phi]|
\geq mc^2\phi^\dagger\phi(1-\frac{1}{2}\zeta^2),
\end{equation}
\begin{equation}
\zeta^2=\frac{\hbar^2}{m^2c^2}\frac{\sqrt{\Delta\phi^\dagger\Delta\phi}}{\sqrt{\phi^\dagger\phi}},
\label{maj0}
\end{equation}
where $\zeta$ is the dimensionless ratio of the Compton length to the characteristic length of the studied non-relativistic problem (e.g. the Bohr radius). Recall that for an electron the Compton length is $\approx 100$ times smaller than the Bohr radius. Thus, the negative contribution to the rest energy is negligible provided that the situation is sufficiently non-relativistic as quantified by $\zeta$. 

Eqs.~(\ref{maj}, \ref{maj0}) are not informative around the zeros of the non-relativistic wave-function, i.e. when $\phi(\bm r)\to 0$, but $\Delta\phi(\bm r)\not\to 0$. Hence let us now assume in (\ref{gaza1}, \ref{gaza2}) that $\phi(\bm r_0)= 0$ and look at $\bm r\approx\bm r_0$. Now 
\BEA
&& -\frac{\hbar^2}{2m}\Re[\phi^\dagger\Delta \phi]={\cal O}(|\bm r-\bm r_0|), \\
\label{val}
&& mc^2(\phi^\dagger\kappa +\kappa^\dagger\phi)={\cal O}(|\bm r-\bm r_0|), \\
&& mc^2\phi^\dagger\phi={\cal O}(c^2|\bm r-\bm r_0|^2), \\
&& mc^2\chi^\dagger\chi={\cal O}(1), 
\label{ordnung}
\EEA
where for (\ref{val}) we assumed that $\kappa(\bm r_0)\not=0$, and where
(\ref{ordnung}) comes from $\bm\nabla\phi(\bm r_0)\not=0$; cf.~(\ref{gla}). Hence the overall energy density $\varrho(\bm r)$ is still positive for $\bm r\approx\bm r_0$. 

The most difficult situation is when $\phi(\bm r_0)=\bm\nabla\phi(\bm r_0)= 0$. 
This case requires specific information about $\kappa$, which we did not attempt to obtain in this paper.
Hence this problem is left open. 

\section{Spin-dependent energy density for Landau levels}
\label{aa5}

So far we discussed the case with zero magnetic field. As usual, the magnetic field elongates space-derivatives subtracting the vector potential times charge, $e\bm A$ from the canonical momentum operator $\bm P=-i\hbar\bm\nabla$ \cite{cohen_laloe,landau}. We confirmed this standard rule via direct calculation. Thus the spin-dependent energy for the magnetic field reads from \eqref{sta}:
    \BEA
    \label{bo}
        &\bm\Upsilon=\frac{\hbar}{4m}\Re\bigg[\phi^\dagger\bm\sigma\times(\bm P + |e|\bm A)\phi\bigg],\\
        &\rho_s=\bm\nabla \bm\Upsilon,
        \label{bodyaga}
    \EEA
where we focus on electron: $e=-|e|$. We emphasize that (\ref{bo}) and (\ref{bodyaga}) are formally gauge-invariant. 

We now turn to calculating (\ref{bodyaga}) for electrons in Landau levels, one of the most important example of magnetic field induced quantum states that is basic for quantum Hall effects, strong correlations, topological states etc; see \cite{tong} for further references and a review. 

We assume a constant, uniform magnetic field $\bm B = [0, 0, B]$ acts on otherwise free electron with charge $-|e|$ and mass $m$. The gauge is fixed as 
\BEA
    \bm A = [-By,0,0]. 
\EEA
Various gauge choices and their differences are discussed in \cite{enk}. 
The stationary Schroedinger-Pauli equation reads
    \BEA
        E\phi &= -\frac{\hbar^2}{2m}\bigg[\bigg(\partial_x - i\frac{m}{\hbar}\omega_By\bigg)^2 + \partial_y^2 + \partial_z^2\bigg]\phi +\nonumber\\&+ \frac{\hbar}{2} \omega_B \sigma_z \phi,\nonumber\\
        \label{brudno}
        \omega_B &= {|e| B}/{m},
    \EEA
    where $\omega_B$ is the cyclotron frequency, and $E$ is the energy. To solve (\ref{brudno}) we take \cite{landau}
    \begin{equation}
        \phi_n(\bm r) \propto e^{i k_x x + i k_z z} \tilde{\phi}_n(y) [s_1, s_2]^T.
        \label{gor}
    \end{equation}
Eq.~(\ref{gor}) is an eigenstate of the canonic momentum $\bm P=-i\hbar\bm\nabla$ 
along $x$ and $z$-axes. Recall that $\bm P$ is not gauge-invariant and hence is not observable.
Eq.~(\ref{gor}) also assumes that the spin state is factorized and amounts to the eigenstate of $\sigma_z$. Eventually, $\phi_n(y)$ is the solution of the 1d harmonic oscillator problem: 
    \BEA
&       \frac{\d^2}{\d y^2} \tilde{\phi}_n(y) + \frac{1}{\hbar^2}\bigg(2mE_n - 2m\hbar s\omega_B-\hbar^2k_z^2\nonumber\\
\label{budu}
&        -m^2\omega_B^2(y-y_0)^2\bigg)\tilde{\phi}_n(y) = 0,\\
&        s =\frac{1}{2} [s_1,s_2]\sigma_z[s_1,s_2]^T = \pm \frac{1}{ 2}, \\
&       y_0 = \frac{\hbar k_x}{m\omega_B}
    \EEA
    where $y_0$ is the center of the harmonic oscillator, and where $[s_1,s_2]^T$ is the eigenvector of $\sigma_z$. Energy levels generated by (\ref{budu}) read (Landau levels):
    \begin{equation}
    \label{harmon}
        E_n = \bigg(n + \frac{1}{2} + s\bigg)\hbar\omega_B + \frac{\hbar^2k_z^2}{2m}.
    \end{equation}
    In these states $\bm \Upsilon = [0, \Upsilon, 0]$ in (\ref{bodyaga}), where
    \BEA
&        \Upsilon = s\frac{\hbar\omega_B}{2}\tilde{\phi}^2_n(y) \bigg[y_0 - y\bigg], \\
&        \rho_s = s\frac{\hbar\omega_B}{2} \bigg[2(y_0 - y)\tilde{\phi}_n(y)\tilde{\phi}_n'(y) - \tilde{\phi}_n^2(y)\bigg].
\label{fri}
    \EEA
It is seen that $\rho_s$ is finite and nullifies together with magnetic field. 

\end{document}